\documentclass{appolb}
\usepackage{epsfig}

\begin{document}
\title{EDMs --- signs of ${CP}$ violation \\ 
  from physics beyond the Standard Model%
\thanks{Presented at the XXXIII International Conference of 
Theoretical Physics, ''Matter to the Deepest'', Ustro\'n, Poland,
September 11-16, 2009 }%
}
\author{Thomas Gajdosik
\address{Vilnius University, 
Sauletekio 9, b. 3, LT-10222 Vilnius, Lithuania \\
and \\
Institute of Physics, 
Savanoriu ave. 231, LT-02300 Vilnius, Lithuania} \vspace{-20pt}
}
\maketitle
\begin{abstract}
The limits placed by the non-measurement of atomic and neutron 
electric dipole moments on $CP$ violating phases beyond the SM 
are found to be not fully justified since the calculations of 
the expected EDMs lack the full understanding 
of the connection between perturbative and nonperturbative regimes
of QCD for the measured boundstates. As a consequence rather 
old subroutines for the evaluation of EDMs are still usable. 
\end{abstract}
  
\section{Introduction}
Electric dipole moments (EDMs) are usually phenomena of extended 
objects, where the charge distribution can be described by a spherical 
multipole expansion. Fundamental pointlike particles can be expected 
to have no dipole moment since they have no internal structure. 
However, this is only true when one ignores the possibility of 
quantum corrections, \ie in the classical limit. Looking closely 
enough at a particle will show the multitude of possible quantum 
corrections, which can and usually will induce an EDM to any 
particle. 

At the level of the Lagrangian or Hamiltonian, an EDM violates 
the invariance of the Lagrangian (Hamiltonian) under parity $P$ and 
under time\-reversal $T$. Assuming $CPT$ invariance this means, that 
the EDM is a $CP$ violating ($CPV$) quantity. In order to generate
$CP$ violation, the Lagrangian has to have complex parameters.

Ignoring the neutrino sector, the perturbative Standard Model (SM) 
has as the single source of $CPV$ the complex phase of the CKM matrix 
$V$. The Jarlskog invariant $J =
 \Im{m}[V_{ij}^{} V_{k\ell}^{} V_{kj}^{*} V_{i\ell}^{*} ]
= s_{1}^{2} s_{2}^{} s_{3}^{} c_{1}^{} c_{2}^{} c_{3}^{} 
\sin\delta_{\mathrm{CP}}$ parametrizes the size of $CPV$ from the 
CKM matrix. As seen in the kaon system, the CPV of the SM is
tiny~\cite{PDG}: $\epsilon \sim 10^{-3}$ and 
$\epsilon^{\prime}/\epsilon \sim 10^{-3}$.

But the SM also includes QCD which allows for an effective CPV 
low energy vacuum angle $\theta$, that is generated as a non 
perturbative effect: 
\begin{equation}
\label{Ltheta}
\mathcal{L}_{\theta} 
:= \mathcal{L}_{\mathrm{dim}=4} 
= \frac{g_{s}^{2}}{32 \pi^{2}} \, \theta \, 
  \varepsilon^{\mu\nu\kappa\lambda} G^{a}_{\mu\nu} G^{a}_{\kappa\lambda} 
\enspace .
\end{equation}
Estimating the contribution of this Lagrangian to the EDM of the neutron 
gives in the most simple model (\cf ~\cite{PR})
\begin{equation}\hspace{-30pt}
  d_{n}(\theta)  
\sim 
  \e \frac{{\theta} m_{*}}{\Lambda_{\mathrm{had}}^{2}} 
\sim 
  {\theta} \cdot ( 6 {\scriptstyle {\,}^{{}_{\times}}} 10^{-17} )
  \e\,\mathrm{cm}
=
  6 {\scriptstyle {\,}^{{}_{\times}}} 10^{9} \cdot {\theta} \cdot 
  d_{n}^{\mathrm{exp}} 
\enspace ,
\end{equation}
which implies $|{\theta}| < 10^{-9}$. This is the strong $CP$ problem, 
which has no generally accepted unique solution yet.

\section{Measurements and calculations of EDMs}
\begin{figure}
\begin{picture}(340,200)(0,0)
\put(0,210){\epsfig{figure=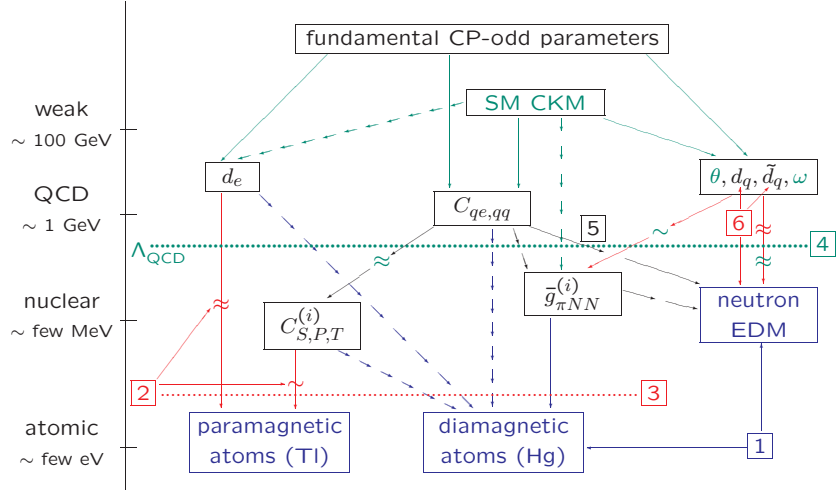,height=340pt,angle=270}} 
\end{picture}
\caption{Sketch of the energy scales involved in calculations of EDMs, 
inspired by~\cite{PR}. The numbers in boxes refer to the developements
after the publishing of PR. 
\label{escales}}
\end{figure}

EDMs are measured today with higher and higher precission: 
the EDM of Thallium~\cite{TlEDM}, 
the EDM of the neutron~\cite{nEDM}, and most recently
the EDM of mercury~\cite{HgEDM}. All of these measurements are done
at very low energies, where QCD is non perturbative. 
The $CPV$ parameters, on the other hand, are defined at high energies, 
where QCD is perturbative. In order to use the measurements to the 
fullest, one has to create a map between the low energy measurements and
the high energy parameters. A map of this kind will necessarily 
involve the discussion of the different energy scales, as started by
Khriplovich et al. 25 years ago. The recent treatment of Pospelov
and Ritz (PR)~\cite{PR}
discusses in particular, how the effective parameters at the low scale 
should be calculated in terms of the $CPV$ parameters at the 
high scale: the calculational method, connecting the parameters 
at the different scales as depicted in Fig~\ref{escales}, should
\begin{itemize}
\item possess chiral invariance, including relevant anomalous Ward 
  identities,
\item account for possible solutions of the strong $CP$ problem,
\item use the same method to calculate all low energy parameters,
\item estimate the precision of the relevant QCD matrix elements,
\item depend only on scale invariant combinations of parameters,
\item give quantitative predictions for the dependence on the different 
  high energy parameters.
\end{itemize}
Unfortunately, such a method does not yet exist. Lattice QCD seems 
to be a good candidate, but does not give definite results yet. 
In lack of such results, PR discuss four approaches to the calculations: 
the non-relativistic $SU(6)$ quark model, naive dimensional analysis, 
chiral techniques, and QCD sum-rules techniques. 

\subsection{Developements after PR}
\begin{enumerate}
\item As predicted by PR, the experimental measurements of the EDMs of 
neutron~\cite{nEDM} and of mercury~\cite{HgEDM} improved significantly 
but without measuring an EDM. So the constraints on the parameters are
still limited by the understanding of the calculated EDMs. 

\item In 2007 Liu {\it et al.}~\cite{Liu:2007zf} analysed atomic EDMs
and reformulated Schiff's theorem~\cite{Schiff-theorem}
at the quantum mechanical operator level. Liu {\it et al.} note, that 
their obtained operator reduces to the usual Schiff moment, that is used 
in the literature to obtain the estimates of the atomic EDMs, only 
under certain additional assumptions. The authors comment also, that 
these assumptions are not generally justified.

That the issue is not settled yet can be seen from the ongoing 
discussion between 
\cite{Sen'kov:2007rt,Zelevinsky:2008pj,Auerbach:2008wa}
and \cite{Liu:2007zf,Liu:2007en}.

\item Already in 2005 Abel and Lebedev~\cite{Abel:2005er} studied 
correlations between the neutron and electron EDMs in the context of 
supersymmetry and remark, that the EDM of Thallium, which is usually 
used to restrict the electron EDM, is dominated by the 
$\theta$-background of the SM, if $\theta > 10^{-15}$. But the 
experimental restriction is only $\theta < 10^{-10}$.

The reported change of $d_{A}^{\mathrm{Schiff}}({}^{199}\mathrm{Hg})$
turns out to be an unresolved typo between the available version 
of~\cite{Latha:2009nq} in the arXiv and the published version, which
was not available at the time of the talk. 

\item \cite{Hernandez:2008db,Hernandez:2008kq} calculate the CPV of 
the effective action of the SM and find, that it is much bigger than 
the Jarlskog invariant. This indicates, that CPV should be treated 
as a non-perturbative effect, which in some way questions the 
perturbative relations between the parameters at the high and the
low scale.

\item This year An, Ji, and Xu~\cite{An:2009zh} give the first 
systematic treatment of the contribution of the four quark operators 
to $d_{n}$ and compare chiral perturbation theory with the results of 
QCD sum-rules. When evaluated with these different descriptions some 
four quark operators give similar, other four quark operators give very 
different contributions to $d_{n}$. 
That result shows a limit of our understanding, how we should describe 
non-perturbative QCD.

\item Already in 2002 Asaga and Fujita~\cite{Asaga:2002hh} noted, 
that the first order contribution of the quarks to $d_{n}$ can 
be shuffled between the EDM and the chromo-EDM of the quarks and 
thus provide an additional suppression of $d_{n}$.

\end{enumerate}
From the arguments {\small \fbox{4}} and {\small \fbox{6}} I draw 
the conclusion, that the perturbative calculations that connect the
high scale parameters $C_{ff}$, $d_{f}$, and $\tilde{d}_{q}$ with 
the lower scale parameters $C_{S,P,T}^{(i)}$, $d_{n}$, and 
$\bar{g}_{\pi NN}^{(i)}$ are not capturing the whole effect, as
they cannot include the non-perturbative nature of QCD. 

\subsection{What we really know}
Though expecting to see EDMs, all measurements up to now could not 
detect any permanent electric dipole moment. These null measurements
of atomic EDMs can be explained by the Schiff's 
screening~\cite{Schiff-theorem}. But the null measurement of the 
neutron EDM still lacks a convincing explanation, which constitutes 
the strong $CP$ problem. 

$CPV$ from the CKM matrix is strongly suppressed by the small mixing 
angles and the expected EDMs generated by the phase of the CKM matrix
are several orders of magnitude smaller than the precision of the 
null measurements. This means that any detection of an EDM points at 
a source beyond the SM.

On the other hand, the null measurements of the EDMs do not restrict
the possible $CPV$ phases in the same way, as the coefficients with 
which these phases contribute to the EDMs are not very well known. 
One obstacle to the exact calculation of these coefficients is that 
QCD is non perturbative in the energy range where the EDMs are measured.

The only EDMs that would not suffer from the QCD uncertainties are the
EDMs of the leptons. A direct measurement of a leptonic EDM 
comes as a by-product from the measurement of the anomalous magnetic 
moment of the muon~\cite{Bailey:1978mn} and gives a limit of 
$d_{\mu} = (3.7 \pm 3.4)\times 10^{-19}\e\,\mathrm{cm}$. There are 
ongoing efforts to improve this measurement directly: 
\cite{Farley:2003wt,Adelmann:2006ab}. 

\subsection{What we should improve}
If we want to use the improving measurements of atomic EDMs we will have 
to use the same low energy Lagrangians and the same corresponding 
Hamiltonians, most probably with the stable particles, i.e. electron, 
proton, and neutron, including their EDMs, as the only available degrees 
of freedom. Since there is quite some freedom of choosing a suitable 
Hamiltonian this step might take the community a while to figure out. 

The next missing piece is the description of the atomic boundstate, 
including the tiny admixtures of the $P$ and $T$ violating interactions, 
that are necessary to desribe a measured EDM. For the electron-nucleus 
part the consensus seems to be growing, but for the nucleon-nucleon 
part the full ab-initio nuclear boundstate calculations are not yet
at the point where the needed heavy nuclei can be calculated reliably. 
This makes the proposed measurement of 
deuterium~\cite{Semertzidis:2003iq,Lebedev:2004va,Liu:2004tq}
a very good alternative to the sole reliance on leptonic EDMs. 

One step further is the understanding of the nucleons as boundstates
of quarks and gluons, which are the degrees of freedom at the high scale.
There are several models that describe the binding of the quarks into 
the nucleons~\cite{Chodos:1974je,Adkins:1983ya,Pirner:1991im}, 
but there is no final agreement~\cite{Lavenda:2006kg} about the calculated 
properties of the nucleon: the different models and the measurements do 
not agree about all properties. This basically means, that we do not yet
fully understand the connection between perturbative QCD, where the high
scale parameters are defined, and nonperturbative QCD, where the 
boundstates are formed. 

One idea might be explored while trying to estimate the effect of EDMs
of the constituents of the bound states: the nucleon as a bound state 
of quarks and gluons and the nucleus as a boundstate of nucleons. The 
argument of the Schiff's screening of EDMs for the atomic boundstates 
operates with the groundstate energy and a displacement operator 
that summarises the EDMs. When the constituents are seen as points, as
the quarks in the nucleons are assumed to be, then the charge and the 
dipole densities are the same and the only violation of the screening 
can come from the relativistic motion of the constituents. In this way 
one could formulate a screening argument similar to \cite{Schiff-theorem} 
for the nucleons and in the following maybe also for the nucleus itself. 

A formulation of such a screening and its evaluation with the 
agreed upon model of the nucleons might also solve the strong $CP$ problem, 
as it gives a symmetry argument for cancellations and hence suppression 
of the estimated EDM of the neutron: the different contributions to the
neutron EDM have to cancel because of the symmetry that describes the 
screening.

\section{Summary and Conclusion}
I failed in the attempt to find significantly improved constraints on 
the $CPV$ phases of the MSSM in the recent literature. Instead I found, 
that the bounds that were published as restrictions should be understood
as the possible reach of experiments that measure atomic or neutron EDMs 
for the detection of $CPV$ phases. 

Null measurements of the EDMs do not restrict the underlying $CPV$
phases in a strict sense, though some phases are more likely small, as 
they would give EDMs that are bigger than the excluded ones, assuming
that there are no cancellations. But these cancellations can be 
explained by an argument similar to Schiff's screening. 

In absence of needed adjustments the subroutines written around 2004 
are still usable to estimate $d_{\ell}$, $d_{n}$, and $d_{\mathrm{Hg}}$ 
in the MSSM. The calculation includes the Weinberg dim=6 operator and 
the 2-loop Barr-Zee type Higgs contributions. $d_{n}$ is calculated with 
two different models for the neutron. These subroutines work as an 
addition to the FeynArts/FormCalc model and driver files for the MSSM.
They can be downloaded from \\
\verb%http://terra.ar.fi.lt/~garfield/EDMs/ %.

\subsubsection*{Acknowledgements}
T.G. thanks the EU Network MRTN-CT-2006-035505 ''HEPTools'' for the 
financial support.

\end{document}